# Coherent satellites in multi-spectral regenerative frequency microcombs

Jinghui Yang[1,*], Shu-Wei Huang[1,2], Zhenda Xie[1,3], Mingbin Yu[4,5], Dim-Lee Kwong[4], Chee Wei Wong[1,*]

[1] Fang Lu Mesoscopic Optics and Quantum Electronics Laboratory, University of California, Los Angeles, CA 90095, USA.

[2] Department of Electrical, Computer & Energy Engineering, University of Colorado, Boulder, CO 80309, USA

[3] School of Electronic Science and Engineering, Nanjing University, Nanjing 210093, China

[4] Institute of Microelectronics, Agency for Science, Technology and Research (A*STAR), Singapore 117865, Singapore

[5] Shanghai Institute of Microsystem and Information Technology, and Shanghai Industrial Technology Research Institute, Shanghai, China

[*] Correspondence to: yangjh@ucla.edu, cheewei.wong@ucla.edu

**Multi-spectral frequency combs provide frontier architectures for laser standoff spectroscopy, optical clockwork, and high-capacity optical communications. Frequency microcombs, aided by their high-quality resonances and inherent third-order nonlinear susceptibility, have demonstrated remarkable impact in frequency metrology and synthesis. However, microcombs are often with limited spectral bandwidth bounded by the intrinsic second-order chromatic dispersion and the consequently low intensities at the spectral edges. To enhance the applications of frequency combs, a spectrally-broad comb generation scheme is often desired. Here we report coherent satellite clusters in multi-spectral regenerative frequency microcombs with enhanced intensities at the octave points and engineered frequency span. Beyond the conventional bandwidth of parametric oscillation, the regenerative satellites are facilitated by higher-order dispersion control allowing for multi-phase-matching in the microcavities. The frequency span of the multi-spectral regenerative satellites is deterministically controlled from 57 THz to 126 THz by pumping at C- and L-bands. We demonstrate that the regenerative satellite preserves the coherence with the central comb through the nonlinear parametric process. We further show the mirrored appearance of the satellite transition dynamics including each comb state that are simultaneously observed at the central comb. These multi-spectral regenerative satellites**

**extend the scope of parametric-based frequency combs and provide a unique platform with wide applications.**

## Introduction

Optical parametric processes serve as the fundamental mechanism for a variety of nonlinear optics phenomena and unique applications involving laser frequency combs[1–3], squeezed state generation[4,5], four-wave mixing with matter waves[6], high-harmonic generation[7], and Bose-Einstein condensation[8]. In the general context of optical parametric oscillation, achieving broadband frequency conversion is usually bounded by chromatic dispersion of the propagating medium and is, however, demanding for frontier applications that require wide coherent spectrum. In a parametric-based frequency microcomb, an overall comb bandwidth of an octave or two-third of an octave will allow for a self-referenced frequency stabilization through *f-to-2f* or *2f-to-3f* carrier-envelope-offset ($f_{CEO}$) technique[9,10], which enables the precise definition of the comb line frequency without requiring an external optical reference. Realizing parametric oscillation in such broad bandwidth without post spectral broadening and with high nonlinear conversion efficiency is comparatively non-trivial in microresonators due to the limited degrees of freedom to control the cavity dispersion. Recently studied dissipative Kerr solitons (DKSs) in microresonators provide an elegant platform for broadband self-referenced combs assisted by dispersive wave[11–15] and have achieved a full octave or a two-third-octave span. This has led to the successful implementation of *2f-3f* self-referencing assisted by external laser sources[10]. This approach, however, still suffers from low comb powers at the octave points for harmonic (*2f* or *3f*) generations and usually requires high-power transfer lasers to overcome the hyperbolic-secant intensity falloff bottleneck of frequency comb[16]. Dispersive wave at both the red- and blue-sides of the pump can be generated with engineered third-order dispersion aided by geometrical design[15,17–19]. Stimulated Raman scattering based processes can amplify the comb intensity on the long-wavelength side effectively[20–22] albeit without an assured frequency conversion on the short-wavelength side. Furthermore multispectral coherent synthesis[23,24] can successfully overcome the power-bandwidth paradox but requires multiple stages of lasers and frequency combs.

Most studied microcombs start with modulation instability (MI) and parametric four-wave mixing (FWM), leading to phase-correlated primary comb modes, followed by sub-comb families generation[21,25–31]. The primarily phase-matched modes, spectrally determined by the local anomalous dispersion, nonlinear frequency shift and pump-resonance detuning, will shape the total bandwidth and general envelope of the overall comb spectrum. These primary modes and their sub-combs are predominantly bounded by the second-order dispersion, striving for a near-zero or normal second-order dispersion for broadband clusters[32,33] and multi-octave parametric oscillation[34]. Quasi-phase-matching can also be introduced to compensate the phase mismatch through periodic dispersion control[35], and recently single-crystalline microcavities have also observed octave-spanned FWM[36,37].

Here we report a different modality in broadband multi-spectral frequency generation, through higher-order dispersion control in silicon nitride microcavities, which achieves regenerative coherent signal and idler satellite structures adjacent to the central frequency comb with higher-order phase-matching in excess of two-thirds of an octave (77 THz). Beyond the conventional phase-matching bandwidth, regenerative satellites are observed in the same microcavity: symmetric satellites with azimuthal cavity mode numbers (in the ≈ $100^{th}$ to $310^{th}$ mode range) symmetric with respect to the coherent pump laser, and dispersive waves with azimuthal cavity mode numbers (in the ≈ $350^{th}$ to $460^{th}$ mode range) asymmetric from the pump, with energy conservation fulfilled in both symmetric and asymmetric instances. Efficient power conversions are observed from the pump to the satellites, on par with that from the pump to the primary intensity lines of the central comb. The symmetric satellite centroid locations match with our theoretical predictions and numerical modeling. With the high-intensity satellites, the overall comb spectra can be in excess of an octave. Furthermore, we examine the dynamical evolution and mutual coherence of these regenerative satellites, via the laser-cavity detuning, through the correlated radio-frequency (RF) amplitude noise spectral densities of the satellites and central comb, and the instrument-limited 113.73-GHz equi-distance spacing across multiple spectral bands of the satellites (O-band) and the central comb (C-band). Via internal modulation instability in each satellite, secondary satellites can be observed through MI process from satellite centers. The multi-phase-matched regenerative satellite combs can serve as promising platforms for ultra-broadband coherent communications, self-referenced frequency combs, and multi-spectral precision sensing and spectroscopy.

## Results

### Frequency microcombs with satellite clusters and dispersive waves

Figure 1 shows a series of measured frequency combs with regenerative satellites under coupled pump power range of 28.5 dBm to 31.5 dBm for different pump laser-cavity detunings. The devices under investigation are silicon nitride microrings with measured loaded quality factor (*Q)* of ≈ 950,000, and ring-waveguide cross-sections of 1600 × 800 $nm^2$ (Figure 1a to 1f) and 1500 × 800 $nm^2$ (Figure 1g). The device fabrication process is detailed in Methods. The inset of Figure 1a shows a scanning electron micrograph of the ring cavity. In this 400 µm diameter ring cavity, the waveguide-cavity coupling is tuned to near critical coupling, with a nearly single mode family of transverse magnetic polarization ($TM_{11}$) across the entire pump wavelength range. The measured free spectral range (FSR) is ≈ 113.9 GHz under cold cavity conditions. On the blue- and red-sides of the central comb ≈ 1425 nm to 1800 nm of Figure 1a to 1f, strong intensity satellite clusters – highlighted in the dashed boxes – are simultaneously observed and even with intensities as high as the central comb. With proper dispersion profile in a slightly different geometry and assisted by dispersive wave generation, the clusters can span close to one full octave or even larger, as shown in Figure 1g.

Consequently, the intracavity phase mismatch per round trip ($\Delta\omega_m$) can be described as $\Delta\omega_m = \omega_m + \omega_{-m} - 2\omega_o$, where the pump frequency is denoted as $\omega_o$, the signal and idler satellites as $\omega_m$ and $\omega_{-m}$ respectively, with $m$ the azimuthal mode number of $TM_{11}$ mode family. Here we calculate the primarily phase matched signal and idler resonance pair, from which the adjacent cascaded comb modes are derived. Conventionally, the central frequency comb is formed by MI bandwidth. The intracavity phase mismatch $\Delta\omega_m$ gradually reaches to zero when the phase shift introduced by the cavity dispersion balances that induced by the intracavity nonlinearity. The case where $\Delta\omega_m$ reaches zero denotes the recent Kerr frequency combs observed to date, which is initiated primarily by parametric modulation instability wherein the second-order dispersion dominates in phase matching, with its corresponding gain bandwidth. We denote this as the central comb for clarity in Figure 1. If the higher even-order dispersion is positive and large enough, however, and able to inversely balance the phase mismatch induced by second-order anomalous dispersion, phase matching ($\Delta\omega_m = 0$) is also possible with a sufficiently large accumulated azimuthal mode number[32]. This results in a phase matching bandwidth in these mode numbers significantly larger than the intrinsic modulation instability gain bandwidth. In the case where the signal and idler cavity mode numbers are symmetric, with equal integer magnitude offset (±) from the pump, and energy conservation of the signal and idler photons are preserved ($\omega_m + \omega_{-m} = 2\omega_0$), we can denote these as the satellites. An analysis of satellite phase-matching conditions is detailed in Methods.

The strong intensity satellites – on the blue- and red-sidebands of the central frequency comb – of Figure 1a corresponds to that of a symmetric satellite comb. In this microcavity, the higher-order dispersion is chosen to be large to support the phase-matching setting. Figure 2a plots the measured group velocity dispersion (GVD) and third-order dispersion (TOD) of the ring cavity, mapped with a Mach-Zehnder-clocked swept-wavelength interferometer (detailed in Methods and Refs. [27]). Figure 2b illustrates the corresponding accumulated cavity phase mismatch of our ring cavity versus mode number. First, at the zero phase match points, the MI-induced phase-matched modes are illustrated in the grey dashed box regions, corresponding also with the central comb structure in Figure 1. Second, with the anomalous dispersion of our ring cavity, the cavity phase match increases initially for increasing mode numbers (up to $m \approx 200$, in this cavity) but folds back towards the zero phase mis-match point again at mode numbers $m \approx 290$. This is illustrated in the black dashed box. Particularly, in this region, symmetric satellites can potentially also be observed. Thirdly, the horizontal red lines denotes the 1×FSR ($m$) and 2×FSR ($m$) phase-matching which can lead to asymmetric paired comb line formation, reminiscent of Faraday instabilities[35,38]. These asymmetric ($m$+1) and asymmetric ($m$+2) cavity phase-matched regions are denoted in Figure 2b, potentially enabled by the inherent large higher-order dispersion of this cavity.

Figure 2c illustrates a zoom-in of the cavity phase-matching around the 250$^{th}$ to 295$^{th}$ azimuthal cavity modes. With the measured FSR, the accumulated dispersion crosses zero at the 291$^{st}$ ± 1 mode for the measurement at 1576.615 nm pump and 30.5 dBm coupled power (blue line). This is the symmetric satellite comb shown in Figure 1a. Adjusting the pumped resonances, Figure 1b and 1c shows the corresponding symmetric satellite spectra for the pump at 1581.881 nm and 1584.704 nm, with 30.5 dBm coupled power. The intensity-weighed centroids of the satellites are illustrated as the magenta and green stars in Figure 2c. Note that from the experiments, the intensity-weighed centroids are close to the primarily phase matched positions as analyzed in the theory above, with difference of less than 3 modes from the intracavity power change due to the detuning adjustment. The effective adjustment of satellites span via pump mode can be understood by perturbation theory applying within the same pumping resonance. One can therefore denote the perturbed frequency of the pump, intensity-weighed centroid frequencies of signal satellite and idler satellite as $\delta\omega_o$, $\delta\omega_m$ and $\delta\omega_{-m}$. Momentum conservation of the satellite centroid frequencies holds as: $\delta\omega_m \cdot GV_m + \delta\omega_{-m} \cdot GV_{-m} = 2\ \delta\omega_o \cdot\ GV_o$, where $GV_i$ represents the group velocity of the $i^{th}$ azimuthal mode. Together with energy conservation $\delta\omega_m + \delta\omega_{-m} = 2\ \delta\omega_o$, this can be readily rewritten as:

$$\frac{\delta\omega_m - \delta\omega_0}{\delta\omega_0} = \frac{2GV_0 - GV_m - GV_{-m}}{GV_m - GV_{-m}} \quad (1)$$

$$\frac{\delta\omega_{-m} - \delta\omega_0}{\delta\omega_0} = \frac{-2GV_0 + GV_m + GV_{-m}}{GV_m - GV_{-m}} \quad (2)$$

The left-hand side of Equation (1) and (2) represent the tunability slope of signal-idler satellite centroids at the given pumping mode $\delta\omega_o$. The phase matched azimuthal mode number $m$ can be calculated as described in Figure 2a, which is related to dispersion profile at $\delta\omega_o$, hence the right-hand side of the equations indicates this tunability is determined by the higher-order dispersion. Plugging into the group-velocity values, the tunability is in the range of 3 to 5 pumping at 1570 nm to 1590 nm for the device majorly investigated in this work, agreeing with the series of measurements on satellite comb spectra (shown in Figure 3). Looking into right-hand side of (1) and (2) further by expanding the group velocity of signal and idler from the pump:

$$GV_m \approx GV_0 + \frac{\beta_{2,0}\Omega}{2} + \frac{\beta_{3,0}\Omega^2}{6} \quad (3)$$

$$GV_{-m} \approx GV_0 - \frac{\beta_{2,0}\Omega}{2} + \frac{\beta_{3,0}\Omega^2}{6} \quad (4)$$

where Ω is the spectral shift, $\Omega = \omega_{\mathrm{m}} - \omega_0 = \omega_0 - \omega_{-m}$, i.e. $\beta_{\mathrm{i},0}$ ($i$ = 2, 3) represent the $i^{th}$ order dispersion terms at pump wavelength, neglecting contribution from higher-order dispersion. Hence Equation (1) and (2) can be expressed as:

$$\frac{\delta\omega_m - \delta\omega_0}{\delta\omega_0} = \frac{\delta\omega_{-m} - \delta\omega_0}{\delta\omega_0} \approx -\frac{\beta_{3,0}\Omega}{3\beta_{2,0}} \quad (5)$$

Therefore, this slope only depends on $\frac{\beta_{3,0}}{\beta_{2,0}}$, given the spectral shift of satellite, the former determined by $\Omega^2 = -\frac{6}{\beta_4}(\beta_2 - \sqrt{\beta_2^2 - \frac{4\beta_4\gamma P}{3}})$ (see Methods for details). Hence by approximation $\frac{\delta\omega_m - \delta\omega_0}{\delta\omega_0} \approx -\frac{\beta_{3,0}\Omega}{3\beta_{2,0}} \approx -\frac{2\beta_3}{\sqrt{-3\beta_2\beta_4}}$ is a function of dispersion terms at pump mode. In optical microresonators, this tunability can be designed by waveguide geometry control. The symmetric satellite comb structure is supported by numerically modeling, with examples for Figure 1a and 1d shown in Supplementary Note 1. Dependence of phase-matching on higher-order dispersion for the symmetric satellites is shown in Supplementary Note 2.

Figure 1d, 1e and 1f show the same microresonator under different pump wavelengths and pump, wherein asymmetric satellites are observed. In Figure 1d, 1e and 1f, we also observed two pairs of comb clusters, on each side of the central comb. These comb clusters are highlighted in the brown, orange and purple boxes for illustration. In Figure 1d the satellites closest to the central comb, at ≈ 1.415 μm and 1.820 μm, are from symmetric phase-matching. The further-spaced higher-order satellites at ≈ 1.303 μm and 1.955 μm are asymmetric. We note that, even with the asymmetric phase-matching extending over two-third of an octave, the peak intensities of the satellites can reach up to -15 dBm – on the same order-of-magnitude intensity as the central comb. Since our output intensity collection is un-optimized over such a large (77 THz) frequency range including cavity-to-waveguide and objective lenses, the intensity of the satellites in the collected Figure 1 spectra should be even higher than -15 dBm. Slightly changed geometric design of the microring lead to a slightly different dispersion profile (simulation comparison shown in Supplementary Note 3), allowing the spanning of satellite microcombs to be an octave (Figure 1g). The high conversion efficiency at the octave points could potentially benefit the *f-2f* self-referenced stabilization application, with greatly reduced power budget.

**Satellite comb structure under different driving conditions**

The primary lines of satellite clusters are generated simultaneously with the primary lines of the central frequency comb and define the fundamental structure of the regenerative frequency combs. This is shown in Figure 3a and 3b, where we plot the comb structures driven at the same resonance mode, but with slightly different detunings at 1581.70 nm (denoted as $\Delta_1$) and 1581.88 nm (denoted as $\Delta_1'$). We note that the satellites grow simultaneously with the primary lines of the central comb, and the primary lines of Figure 3a shape the basic structure of the fully developed satellite and the central combs shown in Figure 3b. Efficient power conversion is achieved from the pump to the satellite sidebands, with the satellite primary intensity close to primary lines of the central comb (Figure 3a) and the fully-defined structure (Figure 3b), even with an un-optimized broadband collection setup. Within the same resonance mode (inset of Figure 3a shows the 1581.70 nm and 1581.88 nm relative detunings), a fully developed satellite can span over 536 modes, equivalently to ≈ 61 THz, simultaneously with a fully developed central comb.

The span of signal-idler symmetric satellite centroids, as analyzed in Equations (1) and (2), can be practically controlled by varying the pumping resonances. This can be seen by zooming into the fully developed signal and idler satellites, as shown in Figure 3c and 3d, under a broad pumping regime from 1570 nm to 1590 nm at 30.0 dBm coupled pump power. We observe and verify that the satellite combs have a symmetric spectral separation of the signal and idler satellites, with respect to each pump mode. This confirms the co-dependence of satellites at blue- and red-sides and the physical basis from signal-idler energy conservation. The magnitude of this signal-idler spectral span is deterministically tuned by the pump wavelength, different from Raman-induced combs[20] which has a phonon-defined vibrational frequency offset. Signal-to-idler satellite centroids span from 57 THz (488 nm) to 80 THz (670 nm) in this resonator. In a different resonator with slightly varied dispersion, assisted by dispersive wave, the span of the satellite centroids can achieve more than one octave, or even a close to one-octave span of ≈ 126.4 THz as shown in Figure 1g, with the short- and long-wavelength dispersive wave peaks reaching 1.2 μm and 2.4 μm respectively. In Figure 3c and 3d, the satellite comb spectra pumped at below 1585 nm wavelengths arise from symmetric family and the spectral shift is observed with a scaling of 3 to 5 satellite azimuthal modes per pump mode, agreeing well with our theoretical estimates in Eqs (1) and (2).

The simultaneous occurrence of both symmetric satellites and dispersive waves can be observed in this microcavity when the pump wavelength is longer than 1585 nm. The dispersive waves are plotted in Figure 3c and 3d as well. Within our ring cavity, the dispersive waves achieve an even larger bandwidth – of 79 THz pumped at 1586.41nm – compared to symmetric satellites. We also note that the intensity-weighted centroid of the satellite can shift within ± 1 mode due to intracavity power changes from detuning adjustment, in a similar manner with the primary modes of the central comb and observed throughout all the pumping modes (e.g. positions of satellite centroids shift by one mode from Figure 3a to 3b). Furthermore, in the central comb when pumped at 1581.88 nm (Figure 3b), the red-side shows more modes and higher intensity, attributable to dispersive wave (DW) due to the large third-order dispersion in our cavity. Examples of the satellite combs with the different DW peaks are detailed in Supplementary Note 4 when pumped at a different spectral region of ≈ 1561.40 nm, where the ratio of second- and third-order dispersion is larger for the same cavity. This DW spectral broadening, however, is still smaller than the symmetric and potentially asymmetric satellite comb spans.

**Observed summary of satellite map versus theoretical comparison**

The multi-phase-matched theory is well-supported by a series of experiments shown in Figure 4a. Under different detunings and on-chip powers (different colored squares), the phase-matched symmetric satellites fall into different spectral locations, agreeing with the theoretical model predictions (black solid curves) over the 140 experimental measurement datasets. Note that, the tunability shown from the simulation is close to the measurement, since this tunability is mainly depending on the ratio of TOD and GVD (from the analysis above on symmetric satellites), whose value is close between simulation and real measurement. Meanwhile, modeling for dispersive waves (green dashed curves) match with observations (different colored triangles) with uncertainty resulting from the uncertainty of higher-order dispersion and refractive index. We also note that both symmetric satellite and dispersive wave can exist under the same pump mode. In Figure 4b and 4c, we illustrate such an example pumping at the same resonance while with a slightly different detuning of 80 pm. This effect is observed in numerical simulations detailed in Supplementary Note 1, where the satellite sidebands can disappear and reappear as the detuning continues to grow.

**Coherence transfer and regeneration with dynamical evolution**

One advantage of the parametric process is the coherence transfer and regeneration. This will potentially lead to a regeneration of dynamical evolution in the multi-spectral structures, in various spectral regimes that are phase-matched. Figure 5a examines the evolution of the signal satellite and their corresponding RF amplitude noise spectral density, with laser-cavity detunings up to 103 pm in the same pump mode, revealing the regenerative evolving dynamics between the signal satellite and the central comb. With pump detuned into the resonance, the satellite first starts with low-noise primary lines (Figure 5a and 5b) with the central comb also in the low-noise state as shown in the right panel of Figure 5a. When the sub-comb lines begin to evolve (stage c), a pristine beat note of 46.0 MHz with its harmonic are observed in the satellite (black curve). Simultaneously, when optically filtering and collecting only the central comb, an instrument-limited identical 46.0 MHz beat note is also observed (green curve). In the central frequency comb, the 46.0 MHz beat note arises from the sub-comb families, which in turn comes from mismatch between the MI-induced phase-matching and local FWM. Our observation of the same 46.0 MHz beat note in the satellite verifies the same underlying mechanism, of the mismatch between MI-induced phase-matching and local FWM, arising in the signal satellite combs. The correlated noise spectrum between the satellite and central comb provides evidence on mutual coherence between central comb and the satellite combs. By detuning the pump deeper into resonance, a self-injection locked state for both the satellite and the central comb is observed. This results in the low-noise coherent state comb (Figure 5d). As the detuning increases further, the satellite becomes broader and transits to high noise states (Figure 5e). The idler satellites, detailed in Supplementary Note 5, also show matching coherence transition and RF evolution as the signal satellites. Besides, our measurements further support a matching coherence evolution of the short-wavelength dispersive wave (substantiated in Supplementary Note 5), in accordance with the symmetric satellites. In the generation of the satellites, coherent satellites with two FSR spacing, high-noise satellites with single FSR spacing, as well as with self-injection locking, can be observed in our measurements through control of the laser-cavity detuning.

The secondary satellites are observed in the symmetric cluster formation, highlighted in the orange dashed boxes of Figure 5a-e. Take the symmetric satellite centered at 1362.5 nm as example (Figure 5a): with the simulated GVD, TOD and FOD of -82.6 $fs^2$/mm, 4.1 $fs^3$/mm and 1806 $fs^4$/mm respectively, the parametric process at 1362.5 nm only supports conventional MI without satellite phase-matching, leading to the formation of secondary satellites. Consequently, the satellites aside from the secondary satellites are generated via the non-degenerate FWM from two central comb frequencies and the satellite centroid, rather than the degeneracy FWM of MI – the former process has higher nonlinear conversion efficiency compared to the latter. This again indicates that the satellite modes, aside from the secondary satellites, are seeded from the central frequency comb and holds spectral correlation.

With these, Figure 5k illustrates the overall parametric process with regards to the satellite formations with coherence transfer and regeneration. Two main simultaneous processes are involved in the satellite evolution: (I) degenerate FWM in forming the conventional MI-induced comb modes, central lines of satellite combs and the secondary satellites when phase matched; followed by (II) satellite evolution generated from non-degenerate FWM from the strong pump, comb modes near the pump and the central line of the satellite. Process (I) defines the fundamental structure of the regenerative frequency comb and process (II) ensures the coherence transfer from the central comb to the satellites, leading to regeneration of comb evolution and comb line spacing between these comb satellites. To further support this framework, we conducted a line-to-line measurement of the regenerative satellite and the central comb across the whole spectrum in a mutually coherent state, determining the 113.730 GHz mode spacing in the O-band at the $5\times10^{-5}$ level ($\approx$ 6 MHz) via a high-precision wavelength meter. We filter each comb mode out with a 1-nm tunable bandpass filter and lock the pump laser to a 1-Hz laser. The wavelength meter has 60-MHz precision, while enables a mode-to-mode FSR spacing precision of 60-MHz/*n* where *n* is the comb mode number away for the pump. This approach also allows the line-to-line measurement at the O-, C- and L-bands from the same instrument. Figure 5l shows several measured comb spacings across the O- and C-band. Both the satellite and the central comb verify an averaged mode-to-mode spacing of 113.730 GHz with an instrument-limited standard deviation of 6 MHz for the satellite (890 kHz for central comb). This further supports a coherence transfer between the satellite and the central frequency comb at the instrument limit, in the existence and formation of the multi-phase-matched satellites.

**Discussion**

In this work we report the formation of coherent satellites in multi-spectral regenerative frequency microcombs based on the higher-order dispersion control, in the azimuthal-mode-number symmetric configurations and assisted by dispersive waves, spanning in excess of two-thirds of an octave. Within the same pump resonance mode, fully developed low-noise signal and idler satellites can span more than two-third octave (77 THz), simultaneously with a fully developed low-noise central comb. Symmetric phase-matching are examined in detail, with a tunability of 3 to 5 satellite modes per pump mode. Coexistence of the symmetric satellite and dispersive wave, along with their exchange, is observed while the signal-to-idler cluster centroids can span more than an octave. Secondary satellites are observed, arising from internal modulation instability at each satellite. Examined robustly over 140 satellite structures, the experimentally observed satellite positions find good match with the theoretical modeling, along with the influence of fourth-order dispersion uncertainties. Mutual coherence between the comb clusters is achieved through non-degenerate four-wave-mixing, validated through the correlated RF beat notes and low-noise power spectral densities in the dynamical evolution, and through the instrument-limited equi-distance spacings between the satellites and the central comb. The studies on multi-spectral and broadband satellites expand the realm of parametric based nonlinear processes, and provide an exceptional chip-scale broadband optical frequency comb source, a suitable platform for self-frequency-referenced oscillator synthesis, and modular precision spectroscopy and sensing in multi-spectral regimes.

**Methods**

**Silicon nitride cavity nanofabrication:** First a 3 μm thick $SiO_2$ layer is deposited via plasma-enhanced chemical vapor deposition (PECVD) on *p*-type 8" silicon wafers to serve as the under-cladding oxide. Then low-pressure chemical vapor deposition (LPCVD) is used to deposit an 800 nm silicon nitride for the ring cavity resonators, with a gas mixture of $SiH_2Cl_2$ and $NH_3$. The resulting silicon nitride layer is patterned by optimized 248 nm deep-ultraviolet lithography and etched down to the buried $SiO_2$ via optimized reactive ion dry etching. Then the silicon nitride cavities are annealed at 1200°C to reduce the N-H overtones absorption at the shorter wavelengths. Finally the silicon nitride cavities are over-cladded with a 3 μm thick $SiO_2$ layer, deposited initially with LPCVD (500 nm) and then with PECVD (2500 nm). The propagation loss of the $Si_3N_4$ waveguide is ≈ 0.2 dB/cm at the pump wavelength.

**Mach-Zehnder-clocked swept-wavelength interferometry:** The cavity transmission was recorded when the laser was swept from 1550 nm to 1630 nm at a tuning speed of 40 nm/s. The sampling clock of the data acquisition is derived from the photodetector monitoring the laser transmission through a fiber Mach-Zehnder interferometer with 40 m unbalanced path lengths, which translates to a 5 MHz optical frequency sampling resolution. Transmission of the hydrogen cyanide gas cell was simultaneously measured and the absorption features were used for absolute wavelength calibrations. Each resonance was fitted with a Lorentzian lineshape to determine the resonance frequency and the quality factor. The cavity dispersion was then calculated by analyzing the wavelength dependence of the free spectral range.

**Analysis on phase-matching conditions for the symmetric satellites:** The parametric gain reaches maxima at the condition as[36,39]: $\sum_{k\geq 1}\frac{\beta_{2k}}{(2k)!}\boldsymbol{\Omega}^{2k}L+2\gamma PL-\delta_0=0$, where $\boldsymbol{\Omega}$ is the sideband frequency shift, $L$ is the cavity length, and $P$ is the intracavity power. Considering up to fourth-order dispersion $\frac{\beta_4}{24}\boldsymbol{\Omega}^4+\frac{\beta_2}{2}\boldsymbol{\Omega}^2+2\gamma P=0$, this leads to the following expression for the spectral shift $\boldsymbol{\Omega}_s$ : $\boldsymbol{\Omega}_s^2=-\frac{6}{\beta_4}\left(\beta_2\pm\sqrt{\beta_2^2-\frac{4\beta_4\gamma P}{3}}\right)$. Different from the previously reported FWM where $\beta_2>0$ and $\beta_4<0$, here in our work $\beta_2<0$ and $\beta_4>0$ and hence intracavity power $P$ must be below $\frac{3\beta_2^2}{4\beta_4\gamma}$. There are two solutions for $\boldsymbol{\Omega}_s^2$ and particularly at the extreme point where $P$ equals $\frac{3\beta_2^2}{4\beta_4\gamma}$ the two solutions equal. We note that the term $\frac{4\beta_4\gamma P}{3}$ cannot be neglected when intracavity power $P$ is sufficiently large due to cavity enhancement. A cavity finesse $F$ of ≈ 600 in this situation corresponds to a power enhancement factor of 95.5. Under mean field and good cavity approximation, this can be simplified to $\boldsymbol{\Omega_s^2 = -\frac{12\beta_2}{\beta_4}}$ [36].

**Dependence of phase-matching on higher-order dispersion:** Taking the first-order partial derivative of $\Omega_s^2$ with respect to $\beta_4$, we get the following expression: $\frac{\partial \Omega_s^2}{\partial \beta_4} = \frac{6\beta_2}{\beta_4^2} \mp \frac{6}{\beta_4^2}\sqrt{\beta_2^2 - \frac{4\gamma P \beta_4}{3}} \pm \frac{4\gamma P}{\beta_4\sqrt{\beta_2^2 - \frac{4\gamma P \beta_4}{3}}}$**.** Under the mean field and good cavity approximation, we can obtain a relationship $\frac{\partial \Omega_s^2}{\partial \beta_4} \approx \frac{12\beta_2}{\beta_4^2}$. This indicates the dependence of spectral shift of the gain on fourth-order dispersion, i.e., decreasing $\beta_4$ increases the spectral shift and vice versa.

**Satellite dynamics measurements:** To perform the satellite comb formation, the high-*Q* microcavity is pumped by a continuous-wave tunable laser followed by an optical amplifier (BkTel THPOA-SL, L-band; IPG EAD-3K-C, C-band), and a polarizer is employed to guarantee the input beam is TM polarized. In our microcavities, the coupling gap is designed to have nearly critical coupling for fundamental TM mode ($TM_{11}$) and weak coupling for the second-order mode ($TM_{21}$) across pump wavelength at 1550 nm to 1620 nm. This ensures the microcavity waveguide can be treated as single-mode operation for TM comb generation. The output comb spectrum is analyzed in both optical domain by optical spectrum analyzers (Yokogawa AQ6375, Advantest Q8384) and RF domain by an electronic spectrum analyzer (Agilent E4402B). Free-spaced filters, WDM filters and tunable O-band filter are used to select the focused O-band, C-band and 2-μm spectral ranges for analysis. Two cascaded O-band WDM fiber filters can realize a suppression ratio of more than 60 dB at C-/L- band, enough to suppress the strong pump and the nearby comb lines in order to analyze the O-band spectra only. A long-pass filter cutoff at 1550 nm and band-pass filters centered at 2000 nm or 2250 nm effectively select idler satellite comb beyond 1900 nm. An InGaAs photodetector (Newport 1611FC-AC, 1 GHz bandwidth, responsive from 900 nm to 1700 nm) is used to measure the amplitude noise of the signal satellite comb and central comb below 1700 nm. An extended InGaAs photodetector (Newport 818-BB-51F, 12.5 GHz bandwidth, responsive from 830 nm to 2150 nm) is used for measuring idler satellite comb.

**Line-to-line measurement of the satellite and central comb:** A high-precision wavelength meter (Bristol-821) is used to measure the wavelengths of the selected frequency lines. Each comb mode is individually filtered out by tunable bandpass filters with 1 nm linewidth (JDSU TB9 covering

C-band and Agiltron FOTF-3-1 covering O-band). The pump laser is phase-locked to a Menlo fiber frequency comb, referenced to ultrastable Fabry-Perot cavity (Stable Laser System), with instantaneous linewidth close to 1 Hz. The schematic setup is shown in Supplementary Note 6. The majority of the measurement imprecision comes from variations of cavity FSR and the pump-cavity detuning. The former leads to the drift of comb spacing and the latter changes comb spacing in a linear manner[40]. With the chip temperature passively stabilized, the variation of cavity FSR is dominated by ambient noise, which can be in the hour time-scale. This measurement approach is bounded only by the precision of the wavelength meter[41]. To reduce the measurement imprecision, we limit the measurement duration to a few minutes.

**Data availability statement:** The datasets generated during and/or analyzed during the current study are available from the corresponding author on reasonable request.

## Acknowledgments

The authors acknowledge discussions with Jinkang Lim, Abhinav Kumar, Hao Liu, Wenting Wang, and Andrey Matsko. The authors acknowledge funding support from the Office of Naval Research (N00014-16-1-2094), the National Science Foundation (awards 1824568, 1810506, 1829071, and Emerging Frontiers in Research and Innovation ACQUIRE 1741707), Lawrence Livermore National Laboratory (contract B622827), the Air Force Office of Scientific Research Young Investigator Award (FA9550-15-1-0081), and the University of California – National Laboratory Office of the President center grant (LFR-17-477237).

## Author contributions

J.Y. designed the devices, performed the measurements, and analyzed the data. Analysis and interpretation of experimental results was conducted by J.Y. and S.W.H. Sample contribution is by M. Y. and D.-L. Kwong. J.Y. and C.W.W wrote the manuscript, with revision contributions by S.W.H. and Z.X.. All authors discussed the manuscript.

## Conflict of interest

The authors declare that they have no conflict of interests.

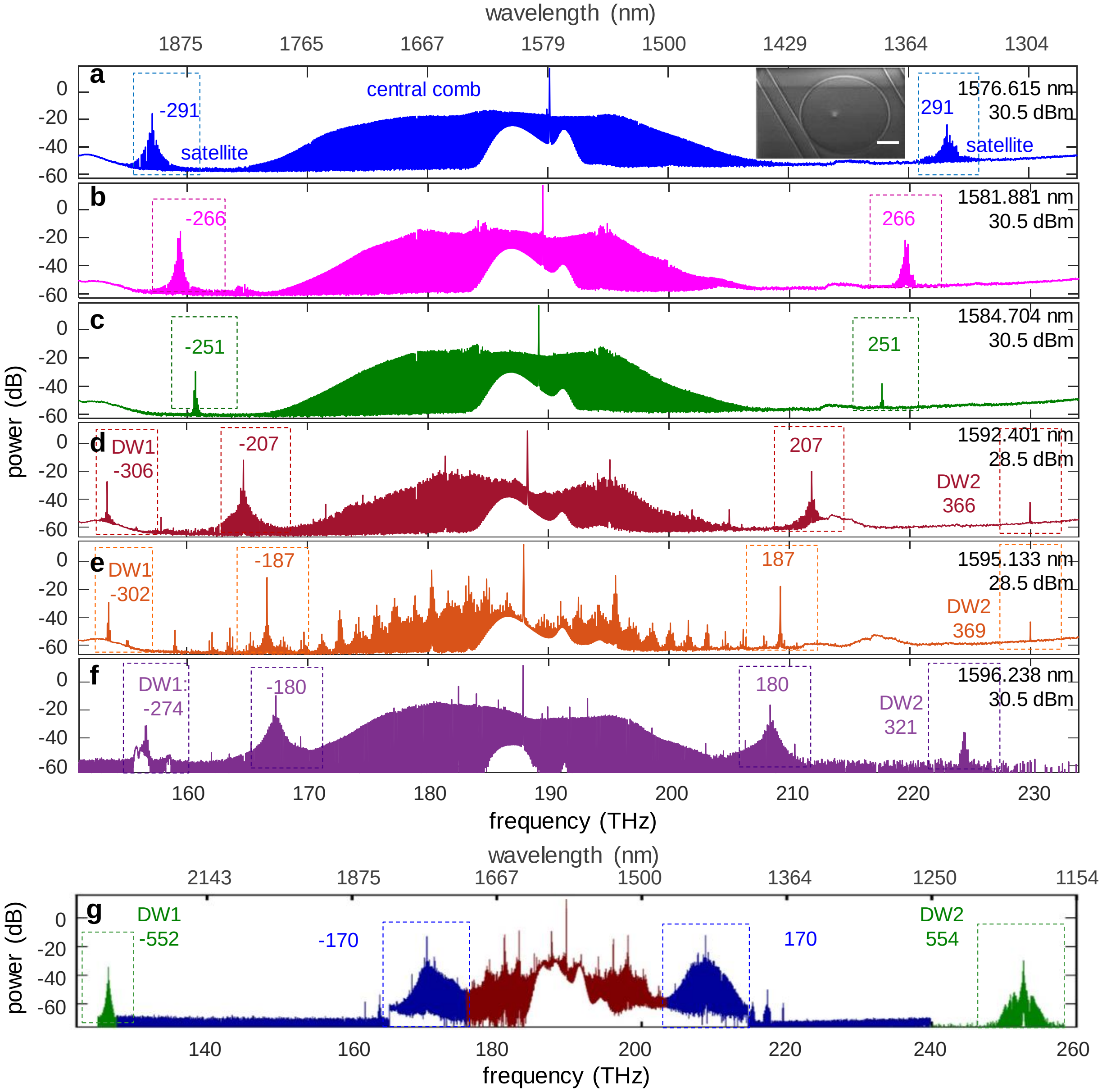

**Figure 1 | Observations of multi-phase-matched satellite frequency combs.** Example spectra of multi-phase-matched frequency comb with both satellite comb and dispersive wave (DW) together with the central comb, pumped at: **a,** 1576.615 nm, **b,** 1581.881 nm, **c,** 1584.704 nm, and **f,** 1596.238 nm with coupled power of 30.5 dBm. Multi-phase-matched satellite comb pumped at: **d,** 1592.401 nm and **e,** 1595.133 nm with coupled power of 28.5 dBm. Two-third octave span with the satellite clusters are achieved in **d** and **e**. **g**, Close-to-one-octave spanning comb pumped at 1581.73 nm, with two satellite centroids locating at 1186.48 nm and 2371.89 nm. The signal-idler satellite centroids overlap within an FSR when the idler satellite is frequency doubled. Inset of **a**, scanning electron micrograph of the microcavity frequency comb, with 400 μm silicon nitride ring diameter and 1600 × 800 nm$^2$ width-height cross-section. Scale bar: 100 μm. The comb mode numbers of the satellites and dispersive waves (DW) are labeled in each panel. The microcavity used in **g** is with slightly different width-height cross-section of 1500 × 800 nm$^2$.

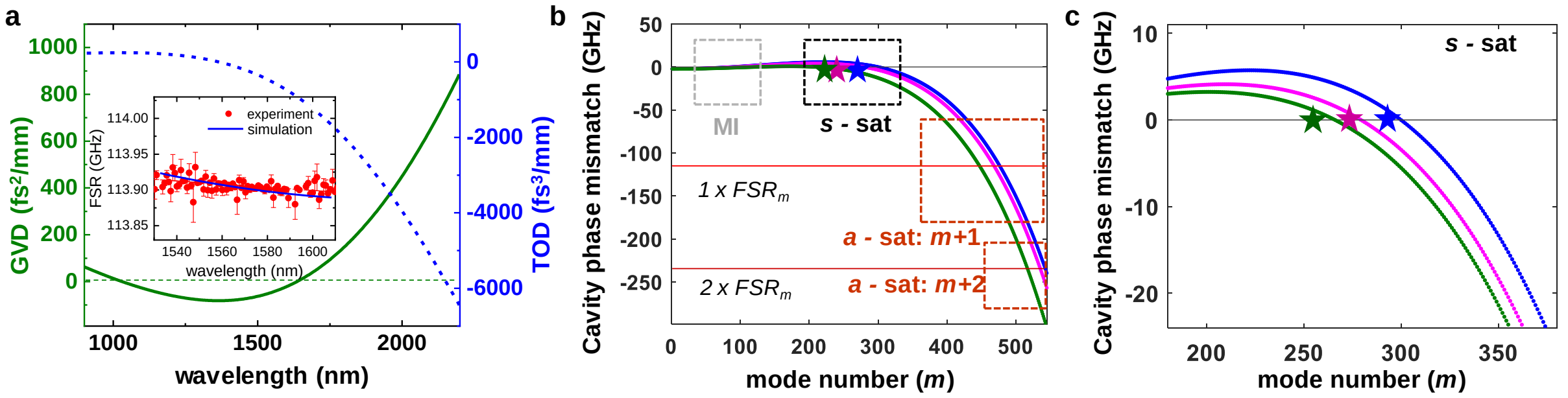

**Figure 2 | Symmetric and asymmetric satellite comb phase-matching: comparison between measurements and modeling. a,** Modeled group velocity dispersion (GVD; green solid curve) and third-order dispersion (TOD; blue dashed curve). Zero dispersion is plotted in green dashed line. Inset: measured GVD by swept wavelength interferometry (red dot) with fitting uncertainty plotted as error bar, along with numerical modeling support (blue solid curve). **b,** Analytical cavity phase mismatch as a function of azimuthal mode number $m$, $\Delta\omega_m = \omega_m + \omega_{-m} - 2\omega_0$, with symmetric satellite comb phase-matching highlighted in black box and asymmetric highlighted in orange box. Highlighted stars are experimentally measured comb number of the satellite centroids pumped at 1576.615 nm (blue), 1581.881 nm (magenta) and 1584.704 nm (green) with coupled powers of 30.5 dBm. Due to the large positive fourth-order dispersion (FOD, $\beta_4$), phase-matching occurs simultaneously at multiple spectral ranges, leading to the satellite comb families at O-band and ≈ 2 μm as shown in Figure 1a to 1c. With the large FOD, the residual dispersion folds backwards, leading to the formation of symmetric satellite combs. If the residual dispersion equals one cavity FSR, parametric oscillation may generate comb mode at (*m*+1), (*m*+2), or even the (*m*+*n*)$^{th}$ cavity mode, leading to the formation of the asymmetric satellite combs. **c,** Zoom-in view of black box in panel **b**. Theoretically calculated phase-matching curves for symmetric satellites, along with the measured satellite centroid position illustrated as datapoints (stars with the same colors).

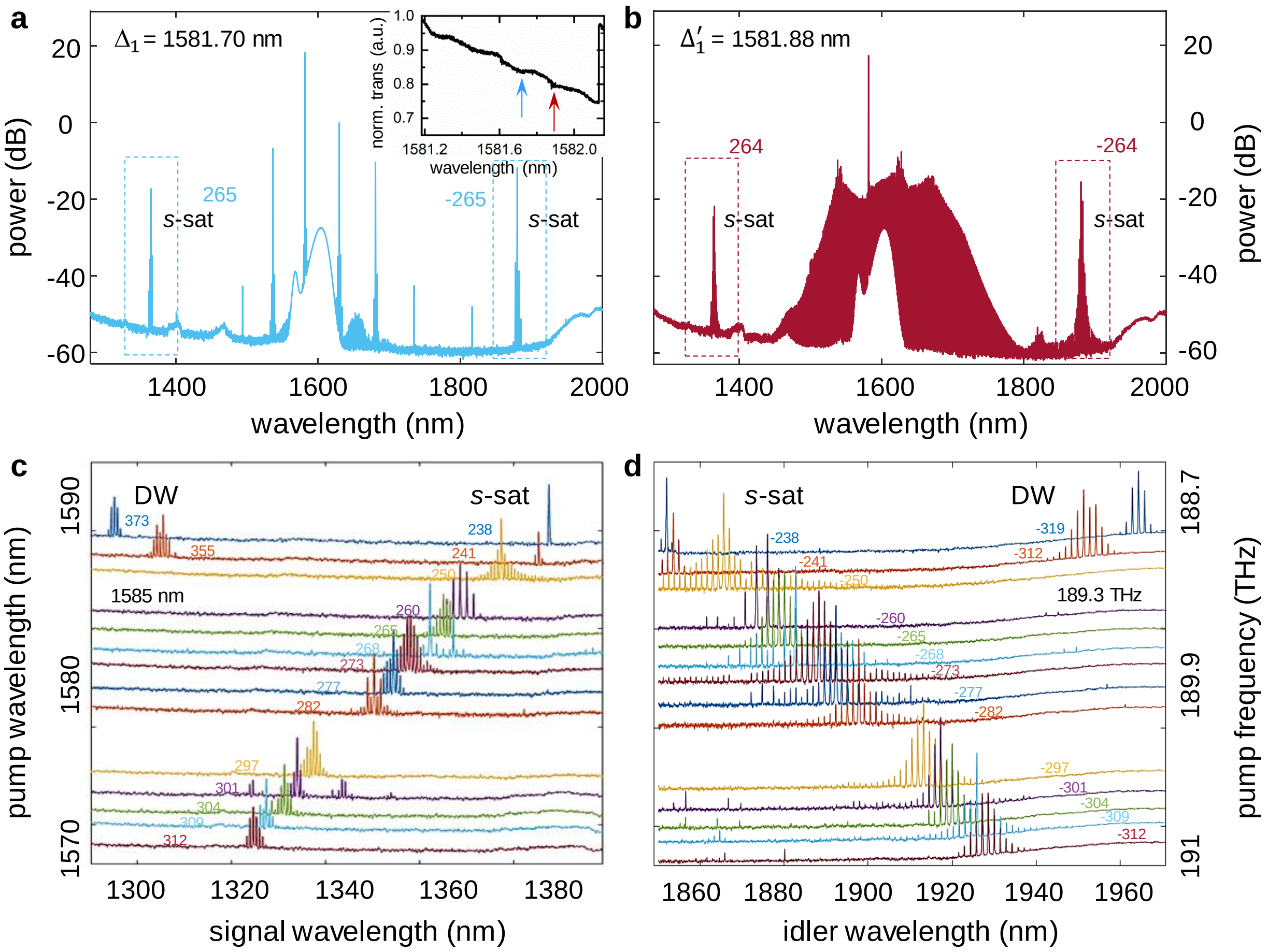

**Figure 3 | Multi-phase-matched satellite combs at ≈ 1350 nm and 1900 nm under different pump conditions. a** and **b,** Zoomed-out frequency comb spectra pumped at the same resonance mode with different detunings ($\Delta_1$ = 1581.70 nm and $\Delta_1'$ = 1581.88 nm; L-band), with coupled power of 30 dBm. Inset of **a**, Pump-cavity transmission as a function of pump wavelength, illustrating the detunings where different satellite comb stages are generated. With more power loaded into the cavity, the intensity-weighed satellite centroid shifts by 1 or 2 cavity modes close to pump due to the increased nonlinear phase shift from higher intracavity power. **c** and **d**, Zoom-in of the ≈ 1350 nm and 1900 nm satellite combs generated when pumped at different resonance modes in the L-band. The vertical axis is the pump wavelength and the horizontal axis is the satellite comb spectra. The comb mode numbers of the satellite centroid are labeled in each panel. Dispersive waves (DW) are observed when pump wavelength larger than 1585 nm.

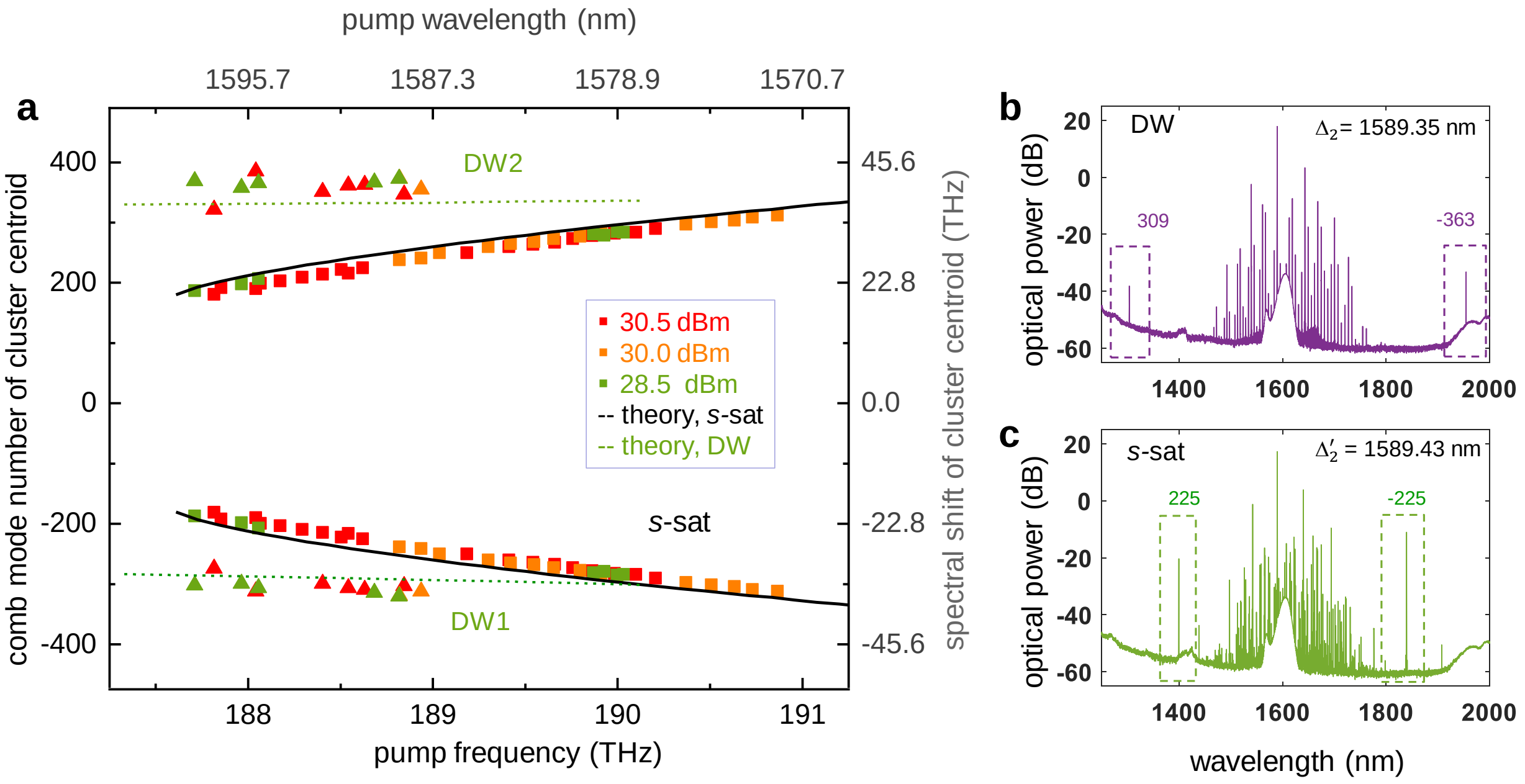

**Figure 4 | Experimental summary map of comb satellites and theoretical comparison. a,** Satellite maps for different pump wavelengths, including parameter sets for three varied coupled pump powers. Under different pump powers and detunings, exchange between symmetric satellite (*s*-sat) and dispersive wave (DW) is observed. Theoretical analysis for symmetric satellites (black solid lines) matches well with our measurements. Modeling for dispersive waves (green dashed lines, DW1 and DW2) match with our measurements with uncertainty resulting from the uncertainty of higher-order dispersion and refractive index. An example of the symmetric satellite (*s*-sat) and dispersive wave (DW) is shown in **b** and **c**, where the resonator is pumped under the same mode but with slightly different detuning. The comb mode numbers of the symmetric satellite and dispersive waves are labeled.

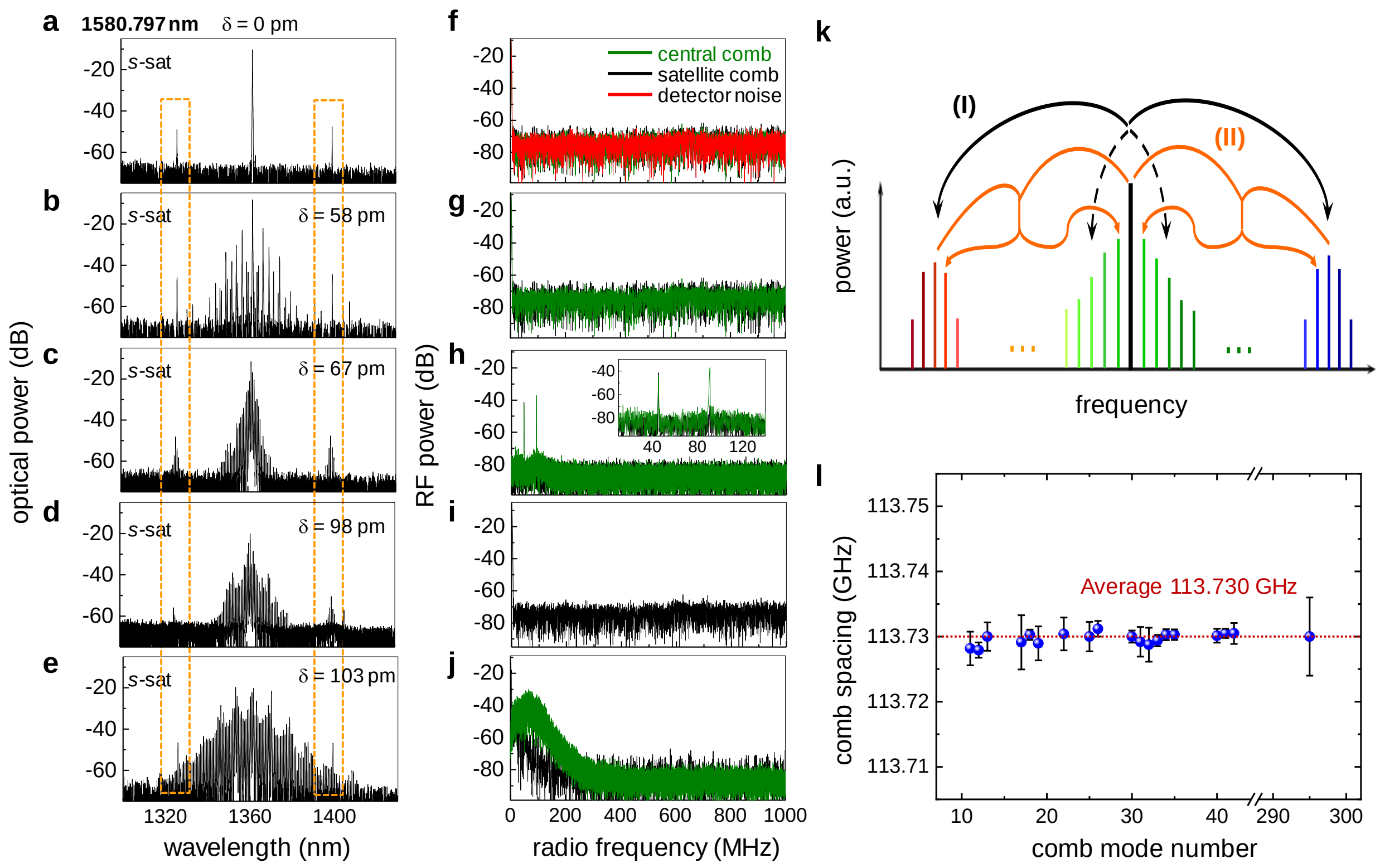

**Figure 5 | Evolution dynamics of the satellites and coherence transfer between the satellites and central frequency comb. a-j, Evolution dynamics of the symmetric satellites. a-e column:** series of satellite spectra with step-wise detuning over 103 pm. The dashed orange boxes denote the secondary satellites within each satellite. **f-j:** corresponding RF amplitude noise measurement. Black and green curves plot for the satellites and the central comb respectively, the latter measured through filtered ≈ 50 nm bandwidth around 1550 nm. Red curve in **f** plots the detector background noise. Both satellite and central comb start with low-noise primary lines as shown in **a, f** and **b, g**. With detuning increases (panel **c**), an instrument-limited identical beat note of ≈ 46.0 MHz can be found between the satellite (black) and central combs (green) as shown in panel **h** . By further detuning the pump, a self-injection locked state is achieved, forming a mutually coherent comb **d**, resulting a measured low amplitude noise as shown in panel **i**. With the detuning further increases, the comb transits to a high noise state as shown in **e** and **j**. **k,** Two consequent processes in satellite formation. (I) degenerate four-wave mixing (FWM) in forming the conventional modulation-instability induced comb modes and central lines of the satellites; and (II) satellite evolution generated from non-degenerate FWM from the strong pump, resonant modes in the vicinity of pump, and the central lines of the satellites; process (II) ensures the coherence transfer from the central comb to the satellites, leading to mutual comb spacing between these comb clusters. **l,** Evidence of coherence transfer. An identical averaged 113.730 GHz comb spacing is experimentally measured across the central comb at C-band and the signal satellite at O-band, with error bars plotted for the standard deviation of comb spacing at each comb mode.